\documentstyle[prb,aps,multicol,graphics]{revtex}
\begin{document}
\draft
\title{Static overscreening and nonlinear response in the Hubbard Model}
\author{Erik Koch}
\address{Max-Planck-Institut f\"ur Festk\"orperforschung,
         Heisenbergstra\ss e 1, 70569 Stuttgart, Germany}
\date{\today}
\maketitle
\begin{abstract}
We investigate the static charge response for the Hubbard model. Using 
the Slave-Boson method in the saddle-point approximation we calculate the
charge susceptibility. We find that RPA works quite well close to 
half-filling, breaking, of course, down close to the Mott transition. 
Away from half filling RPA is much less reliable: Already for very
small values of the Hubbard interaction $U$, the linear response
becomes much more efficient than RPA, eventually leading to overscreening 
already beyond quite moderate values of $U$. To understand this
behavior we give a simple argument, which implies that the response
to an external perturbation at large $U$ should actually be strongly
non-linear. This prediction is confirmed by the results of exact 
diagonalization.
\end{abstract}

\begin{multicols}{2}
\section{Introduction}
Motivated by the surprising accuracy of the random phase approximation (RPA)
for a half-filled, generalized Hubbard model with orbital 
degeneracy,\cite{screen} we ask how well RPA reproduces the screening in
correlated systems in general. While it properly describes the screening 
when the kinetic energy is much larger than the interaction energy, RPA becomes
wrong in the opposite limit. Its main deficiency is that it completely misses
the break-down of the screening at the Mott transition. Nevertheless, as
the quantum Monte Carlo calculations in Ref.\ \onlinecite{screen} have
shown, RPA gives a surprisingly accurate description of the static charge
response on the metallic side of the Mott transition, until the system is
quite close to the transition point. Away from half filling (or, more 
generally, integer filling) one would expect the system to become more 
metallic, and hence, RPA to work even better. We will see, however, that this 
not the case.

In the present study we use a Slave-Boson 
method\cite{KotliarRuckenstein,woelfle} to calculate the linear
charge response. The advantage over more elaborate methods like quantum Monte
Carlo (QMC) or exact diagonalization are (i) the possibility to treat very 
large systems, such that finite size effects become negligible, 
(ii) the efficiency of the method, which allows us to thoroughly study the
whole parameter space of different densities and interactions, and
(iii) the possibility to directly calculate the linear response, which in
Lanczos or QMC calculations has to be extrapolated from several calculations
for different perturbations of finite strength.

There have been already a number of works determining the linear response
from a one-loop expansion around the Slave-Boson saddle-point 
solution,\cite{rasul,lavagna,woelfle2,li1,li2,fresard} but they mainly 
focussed on the structure factors and the spin response. In contrast, here 
we are interested in the static density response, which can be calculated at 
the saddle-point level, avoiding the problems encountered when considering 
Gaussian fluctuations of the Slave Bosons.\cite{jolicoeur,arrigoni1,arrigoni2} 
In that sense our approach is related to that of Refs.\ \onlinecite{seibold}
and \onlinecite{ziegler}, with the main difference that we consider the 
linearized saddle-point equations, in order to directly obtain the linear 
response.
As for the accuracy of the results, based on the experience of previous
works\cite{li2,lilly,preuss,zimmermann} we expect that the static response
should be well described.

The model we consider here is the one-band Hubbard model with nearest and
next-nearest neighbor hopping on a square lattice. It is introduced in 
Sec.\ II.A. We are in particular interested in the response to a point charge. 
In Sec.\ II.B we describe the approach for calculating the charge response
using Slave-Bosons at the saddle-point level. To check the method and its
accuracy, we compare with the result of exact diagonalization. In Sec.\ II.C
we give the results of our calculations: the screening as a function of
filling and interaction. The most striking result is that, already
for quite moderate interaction, the systems shows a response
that is stronger than the perturbation (overscreening). The response,
of course, strongly depends on the doping and changes with the next-nearest
neighbor hopping, being enhanced close to the van Hove singularity. 
Comparing with the random phase approximation, we find that, contrary to
our expectation, RPA works best close to half filling.
In Sec.\ III we give an interpretation of these results. Both, doping
dependence and overscreening can be understood using simple large-$U$
arguments. We also discuss why, by the same arguments, we do not obtain
overscreening in the RPA and that overscreening does not mean that
the systems becomes unstable. The argument for explaining the overscreening
does, however, imply that the charge response of the non-half-filled Hubbard 
model should be strongly nonlinear, which we numerically confirm using exact 
diagonalization. Indeed, we find that the stronger the response, the more
nonlinear it will be.
A conclusion, Sec.\ IV, closes the presentation.

We finally would like to point out that overscreening
in the Hubbard model has been observed before and a proof, although in
quite a different spirit from ours, for the necessity of overscreening 
for large interactions has been given.\cite{schuettler}

\section{Slave-Boson calculations}
\subsection{Model}
We consider the one-band Hubbard model with nearest and next-nearest neighbor 
hopping ($t$ and $t'$, respectively)
\begin{displaymath}
 H=-t   \!\!\sum_{\langle i,j\rangle,\sigma}\!
        f_{j,\sigma}^\dagger f_{i,\sigma}^{\phantom{\dagger}}
   -t'\!\!\!\sum_{\langle\langle i,j\rangle\rangle,\sigma}\!\!
        f_{j,\sigma}^\dagger f_{i,\sigma}^{\phantom{\dagger}}
   + U \sum_i n_{i,\uparrow} n_{i,\downarrow}
\end{displaymath}
on a square lattice with lattice constant $a$. For the noninteracting system
($U=0$) the dispersion relation is
\begin{displaymath}
 \varepsilon_k=-2t(\cos(k_xa)+\cos(k_ya))-4t'\cos(k_xa)\cos(k_ya)
\end{displaymath}
and the density of states has a logarithmic van Hove singularity at $4t'$.

We ask for the response to the perturbation by an external point charge $c$ 
at site $r_c$, corresponding to an extra term $c\,U \sum_\sigma n_{c,\sigma}$
in the Hamiltonian. Expanding the external potential in plane waves,
we find, given the static susceptibility $\chi_q$ for wave vector $q$, for 
the induced charge at site $r_i$
\begin{equation}\label{respc}
 \delta n_i = \sum_{q\ne0} \chi_q\,\delta V_q 
            = {c\,U\over N}\sum_{q\ne0} \chi_q\;e^{iq(r_i-r_c)} ,
\end{equation}
where we have excluded the $q=0$ component, since we are really interested
in the response for a system with fixed number of electrons. Below we will
particularly focus on the response at the site with the point charge
$\Delta n \equiv \delta n_{i=c}$.

For the noninteracting system the susceptibility (per spin) $\chi_q^{(0)}$ is 
given by (we give a more general definition, which we will need below)
\begin{equation}\label{susc0n}
 \chi_q^{(0,n)} = {1\over N}\sum_k
  {\varepsilon_k^n f_k - \varepsilon_{k+q}^n f_{k+q} \over
   \varepsilon_k       - \varepsilon_{k+q}}
\end{equation}
for $n=0$. Here $N$ is the number of lattice sites, the sum is over all states 
(first Brillouin zone) and $f_k=(1+\exp(-\beta(\varepsilon_k-\mu)))^{-1}$
is the Fermi-Dirac function. In case of degeneracy the quotient in 
(\ref{susc0n}) becomes the differential. In RPA the susceptibility is
\begin{displaymath}
 \chi_q^{RPA} = {2\chi_q^{(0)}\over1-2U\chi_q^{(0)}} .
\end{displaymath}

\subsection{Method}
As an efficient way for calculating the static response we choose the
Slave-Boson method, which will allow us to easily explore the parameter
space (fillings $n_\sigma$, interactions $U$, hoppings $t'$) for quite big
lattices ($100\times100$) and low temperatures ($\beta=100/t$). We use the
formalism developed by Kotliar and Ruckenstein\cite{KotliarRuckenstein}
(since we are only interested in the density response of the paramagnet, we 
need not explicitly take care of the spin-rotation invariance).\cite{woelfle}
Instead of studying Gaussian fluctuations of the Slave-Boson action around
the saddle-point (one-loop expansion)\cite{rasul,lavagna,woelfle2,li1,li2}, 
we stay at the mean-field level, simply linearizing the saddle-point equations
for the perturbed system. In that sense, our approach is closer to that
of Refs.\ \onlinecite{seibold} and \onlinecite{ziegler}, where the 
full saddle-point equations for a system with finite perturbation were solved. 
We thus avoid the problems related to the representation of the hopping 
operator.\cite{jolicoeur,arrigoni1,arrigoni2}
The main advantage of our linearized saddle-point equations is, that for a 
plane-wave perturbation the resulting linear system becomes block diagonal, 
with identical blocks of size $5\times5$, independent of system size.
We expect our method to give a good description of the true screening,
given the good experience, in particular for static 
quantities.\cite{li2,lilly,preuss,zimmermann}

\end{multicols}
We now give a brief outline of the linearization procedure. We start from
the Kotliar-Ruckenstein formulation of the Slave-Boson 
method.\cite{KotliarRuckenstein} It maps the physical Fermions onto composite
particles of pseudofermions and four Bosons, representing empty ($e_i$), singly
($p_{i\sigma}$), and doubly ($d_i$) occupied sites. Constraints, ensuring 
consistency between pseudofermions and Bosons, enforced by Lagrange parameters
($\lambda_i^{(1)}$ and $\lambda_{i\sigma}^{(2)}$), are introduced to eliminate 
the unphysical states in the enlarged Hilbert space. In this enlarged space, 
the Hamiltonian can be represented in a form, having kinetic terms with mixed
fermionic-bosonic character (renormalized hopping), while the
interaction becomes purely bosonic and bilinear. The pseudofermions,
merely occurring in bilinear terms, can then be integrated out exactly, and
one is left with a purely bosonic action integral (including the Lagrange
parameters that ensure the coupling to the pseudofermions). The simplest
approach to the bosonic action integral is the saddle-point approximation,
in which all Bose fields are time independent. The Slave-Boson action then
takes the form 
\begin{displaymath}
 \int\limits_0^\beta d\tau\,S(\tau)=
  \beta\sum_i\left[
   \lambda_i^{(1)}e_i^2
  +\sum_\sigma\left(\lambda_i^{(1)}-\lambda_{i\sigma}^{(2)}\right)p_{i\sigma}^2
  +\left(U+\lambda_i^{(1)}-\sum_\sigma\lambda_{i\sigma}^{(2)}\right)d_i^2
  -\lambda_i^{(1)}\right]
 -\sum_{n\sigma}\ln\left[1+e^{-\beta\varepsilon_{n\sigma}}\right]
\end{displaymath}
where
\begin{displaymath}
 z_{i\sigma}={e_i p_{i\sigma}+p_{i-\sigma} d_i \over
       \sqrt{1-d_i^2-p_{i \sigma}^2}\;\sqrt{1-e_i^2-p_{i-\sigma}^2}}
\end{displaymath}
and the $\varepsilon_{n,\sigma}$ are the eigenvalues of the renormalized
Hamiltonian 
\begin{equation}\label{renHamil}
 H_\sigma=-\sum t_{i,j}z_{j,\sigma}z_{i,\sigma}\,
            f_{j,\sigma}^\dagger f_{i,\sigma}^{\phantom{\dagger}}
         + \sum_i \left(\lambda_i^{(2)}-\mu+V_i\right)\,n_i .
\end{equation}
The mean-field values of the Bosons and the Lagrange parameters are determined
from the saddle-point equations
\begin{displaymath}
 \begin{array}{lcr@{\,}ccl}
 \displaystyle 0=\beta^{-1}\,{\partial S\over\partial e_i}&=&
  2\,\lambda_i^{(1)}\;\;\;&e_i&+&
  {\displaystyle\sum_{n\sigma'}}\,{1\over1+e^{-\beta\varepsilon_{n\sigma'}}}\;
     {\partial\varepsilon_{n\sigma'}\over\partial e_i}\\[2ex]
 \displaystyle 0=\beta^{-1}\,{\partial S\over\partial p_{i\sigma}}&=&
  2\left(\lambda_i^{(1)}-\lambda_{i\sigma}^{(2)}\right)&p_{i\sigma}&+&
  {\displaystyle\sum_{n\sigma'}}\,{1\over1+e^{-\beta\varepsilon_{n\sigma'}}}\;
    {\partial\varepsilon_{n\sigma'}\over\partial p_{i\sigma}}\\[2ex]
 \displaystyle 0=\beta^{-1}\,{\partial S\over\partial d_i}&=&
  2\left(U+\lambda_i^{(1)}-{\displaystyle\sum_\sigma}\,
         \lambda_{i\sigma}^{(2)}\right)&d_i&+&
  {\displaystyle\sum_{n\sigma'}}\,{1\over1+e^{-\beta\varepsilon_{n\sigma'}}}\;
    {\partial\varepsilon_{n\sigma'}\over\partial d_i}\\[4ex]
 \displaystyle 0=\beta^{-1}\,{\partial S\over\partial \lambda^{(1)}_i}&=&
  \multicolumn{2}{c}{e_i^2+{\displaystyle\sum_\sigma}\,p_{i\sigma}^2+d_i^2-1}\\[2ex]
 \displaystyle 0=\beta^{-1}\,{\partial S\over\partial\lambda^{(2)}_{i\sigma}}&=&
  \multicolumn{2}{c}{-p_{i\sigma}^2-d_i^2}&+&
  {\displaystyle\sum_{n\sigma'}}\,{1\over1+e^{-\beta\varepsilon_{n\sigma'}}}\;
    {\partial\varepsilon_{n\sigma'}\over\partial \lambda_{i\sigma}}\\
 \end{array} 
\end{displaymath}
For finite temperature $\beta$, this nonlinear
system of equations has to be solved self-consistently, since the eigenvalues 
$\varepsilon_{n,\sigma}$ of the renormalized Hamiltonian (\ref{renHamil}) 
depend on the mean-field values of the bosonic fields. The Fermionic terms 
are most easily evaluated in the form 
\begin{displaymath}
 \sum_{n,\sigma'} f(\varepsilon_{n,\sigma'})\,
   {\partial\varepsilon_{n,\sigma'}\over\partial x_\sigma}
 = -{1\over\pi}\Im\int_{-\infty}^\infty d\epsilon\,f(\epsilon)\,
     \mathrm{Tr}\left(G_{\sigma'}(\epsilon)\,
                      {\partial H_{\sigma'}\over\partial x_\sigma}\right) ,
\end{displaymath}
where $G_\sigma(\epsilon)=(\epsilon-H_\sigma)^{-1}$ is the Greens function.
For the homogeneous system, the Boson variables are
independent of the site index, and solving the saddle point equations 
for the paramagnet ($p\equiv p_\uparrow=p_\downarrow$) essentially reduces 
to self-consistently solving a 3rd-order polynomial in $d^2$.

For calculating the linear response, it is easiest to consider external 
potentials $V_i=\delta V_q\,\cos(q\,r_i)$, which give rise to a response: 
$n_i=n+\delta n\,\cos(q\,r_i)$ and 
$e_i^2=e^2+2e\,\delta e\,\cos(q\,r_i)$, \dots.
Expanding to first order and using the relation
\begin{displaymath}
 -{1\over\pi}\Im\int_{-\infty}^\infty d\epsilon\,f(\epsilon)\,\epsilon^n
    \sum G_{i,j}(\epsilon)\,\cos(q\,r_j)\,G_{j,i}(\epsilon)
 = \chi_q^{SBMF,n}\,\cos(q\,r_i) ,
\end{displaymath}
where $\chi_q^{SBMF,n}$ is calculated as (\ref{susc0n}), but using the 
eigenvalues of the renormalized Hamiltonian (\ref{renHamil}) for the 
homogeneous solution, we can factor out the coordinate dependence; 
i.e.\ we find that the linearized the saddle-point equations are block 
diagonal, with identical $5\times5$ blocks of the form
\begin{displaymath}
 \left\lgroup
 \begin{array}{lllrl}
  A z_e z_e + D( z_{ee}-{z_e\over e}) &
  A z_e z_p + D\;z_{ep}               &
  A z_e z_d + D\;z_{ed}               &  e & B z_e    \\
  A z_p z_e + D\;z_{pe}               &
  A z_p z_p + D( z_{pp}-{z_p\over p}) &
  A z_p z_d + D\;z_{pd}               & 2p & B z_p-2p \\
  A z_d z_e + D\;z_{de}               &
  A z_d z_p + D\;z_{dp}               &
  A z_d z_d + D( z_{dd}-{z_d\over d}) &  d & B z_d-2d \\[2ex]
  e     &       2p &        d & 0 & 0 \\
  B z_e & B z_p-2p & B z_d-2d & 0 & C
 \end{array}
 \right\rgroup
 \left\lgroup
  \begin{array}{c}
   de/dV_q\\ dp/dV_q\\ dd/dV_q\\[2ex] d\lambda^{(1)}/dV_q\\ d\lambda^{(2)}/dV_q
  \end{array}
 \right\rgroup
 =
 \left\lgroup
  \begin{array}{l}
   -B z_e\\ -B z_p\\ -B z_d\\[2ex] \quad0\\ -C
  \end{array}
 \right\rgroup
\end{displaymath}
Here the $z_e$, $z_{ee}$, \ldots denote the partial derivatives of $z$, and, 
since we have set $p\equiv p_\sigma$ (paramagnetic solution), the partial 
derivative with respect to $p$ means the sum of the partial derivatives with 
respect to $p_\uparrow$ and $p_\downarrow$. 
Also, we have introduced
$A=4 \chi_q^{(0,2)}-6 \epsilon_0 $, 
$B=2(\chi_q^{(0,1)}-n_\sigma)/z  $, 
$C=  \chi_q^{(0,0)}          /z^2$, and
$D=2z \epsilon_0$, where $\epsilon_0$ is the energy density per spin and
the $\chi_q^{(0,n)}$ the susceptibilities (\ref{susc0n}) for the
noninteracting part of the unrenormalized Hamiltonian taken at the renormalized 
temperature $z^2\beta$ and Fermi energy $(\mu-\lambda^{(2)})/z^2$.
Solving the linear system, the susceptibility is given by 
$\chi_q^{SBMF}=dn_q/dV_q=4(p\,dp/dV_q + d\,dd/dV_q)$, and the response
to a point charge follows from eqn.\ (\ref{respc}).
\begin{multicols}{2}

\begin{figure}
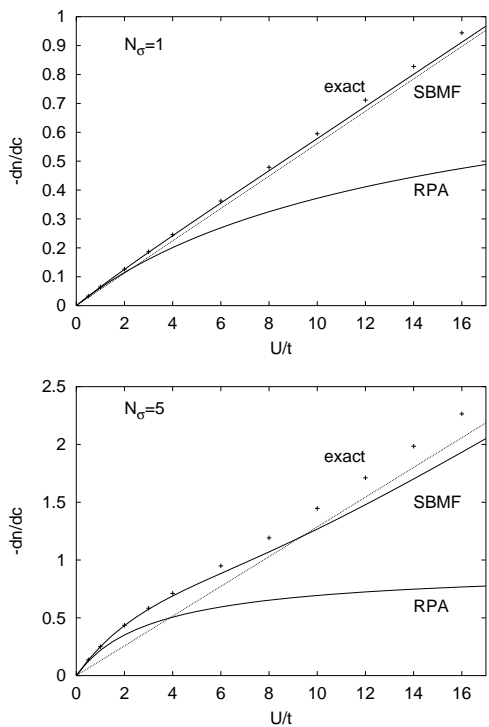

\centerline{\resizebox{2.5in}{!}{\rotatebox{270}{\includegraphics{lanc01.epsi}}}}
\vspace{2ex}
\centerline{\resizebox{2.5in}{!}{\rotatebox{270}{\includegraphics{lanc05.epsi}}}}
 \vspace{2ex}
 \caption[]{\label{lanccheck}
  Comparison of the linear response calculated by the Slave-Boson method with
  exact diagonalization for a $4\times4$ Hubbard model ($t'=0$) with
  $N_\sigma$=1, 5 electrons of each spin (closed shells). The dotted lines
  show the leading order of the asymptotic expansion of the Slave-Boson result. 
  RPA is given for comparison.
 }
\end{figure}

To check the method, and to verify our code, we compare the results
of the Slave-Boson calculation to the linear response calculated by
exact diagonalization for a $4\times4$ Hubbard model. As can be seen
from Fig.\ \ref{lanccheck}, the Slave-Boson results basically reproduce
the response, although they tend to slightly underestimate the screening.
The good agreement does not come as a 
surprise.\cite{li2,lilly,preuss,zimmermann}

While the explicit expressions for the susceptibility $\chi_q^{SBMF}$ are 
somewhat messy, the results of an asymptotic expansion for large $U/t$ are 
fairly 
compact. We have to distinguish two cases: For $n_\sigma<1/2$, to leading order
\begin{displaymath}
 \chi_q^{SBMF}\sim\! {-2(1-n_\sigma)^3(1-2n_\sigma)\chi_q^{(0)}\over
          (\chi_q^{(2)}\!-4(1-n_\sigma)\epsilon_0)\chi_q^{(0)}\!
          -(\chi_1-1+2(1-n_\sigma)^2)^2} .
\end{displaymath}
For $n_\sigma>1/2$ we have
\begin{displaymath}
 \chi_q^{SBMF}\sim {-2n_\sigma^3(2n_\sigma-1)\chi_q^{(0)}\over
         (\chi_q^{(2)}-4n_\sigma\epsilon_0)\chi_q^{(0)}
         -(\chi_q^{(1)}-2n_\sigma^2)^2} .
\end{displaymath}
A superficial look at the two expressions might suggest that, while they are
almost symmetric under the substitution $n_\sigma\leftrightarrow(1-n_\sigma)$, 
they violate particle-hole symmetry. This is, however, not the 
case. For $t'=0$, the results are particle-hole symmetric, due to the 
properties of the generalized susceptibilities.
To give an impression of how the asymptotic expansion works for moderate
values of $U$, we included it in Fig.\ \ref{lanccheck}.

The important conclusion to draw from the asymtotic expansion is that for
large $U$, to leading order, the susceptibility is independent of the
interaction. Therefore, as long as the large-$U$ value does not happen to
vanish (as is the case for $n_\sigma=0,\,1/2,\,1$), the linear response to a
point charge $dn/dc$ will, to leading order, grow linearly with $U$ --- 
beyond any limit.

\subsection{Results}
\subsubsection{Overscreening}
The results of the Slave-Boson mean-field calculations for the response to a 
point charge as a function of filling $n_\sigma$ and interaction $U$ for
different hoppings $t'$ are shown in Figure \ref{sbmfresponse}. The most 
striking feature is that only exactly at half-filling, the response vanishes 
(at the Mott transition), while for any other filling $n_\sigma\ne1/2$ the 
induced charge density on the site with the test charge eventually increases 
linearly with $U$. The reason is that, in the limit of $U$ going to infinity, 
the susceptibility of the non-half-filled system stays finite, only for
$n_\sigma=1/2$ does $\chi_q$ vanish (see Fig.\ \ref{largeUchi}). This 
necessarily means that for strong 
enough interaction and $n_\sigma\ne1/2$ the induced charge will be larger than 
the perturbation. What is quite surprising is, that this overscreening already 
sets in for very moderate values of $U$. As a function of the next-nearest 
neighbor hopping, we find overscreening for (see the first contour line in
Fig.\ \ref{sbmfresponse}):
\begin{displaymath}
 \begin{array}{l|ccccc}
  t'/t            & 0 & -0.2 & -0.3 & -0.4 & -0.5 \\ \hline
  U/t\gtrsim\quad & 6 &  5   &  4   &  3   &  2
 \end{array}
\end{displaymath}
We also remark that, according to the comparison with the exact results for
a small system (Fig.\ \ref{lanccheck}), the Slave-Boson results might even
somewhat underestimate the true response.

The second striking feature is the strong doping dependence of the response.
It tends to be strongest for quarter filling ($n_\sigma=1/4$ or $3/4$),
while it vanishes when approaching the Mott- or band-insulating regions. 
This doping dependence is somewhat modified by the next-neighbor hopping $t'$: 
The screening becomes enhanced in the neighborhood of the van Hove singularity 
(at $4t'$). For $t'/t=-1/2$, where the van Hove singularity is at the lower
band-edge, this enhancement is particularly pronounced, while for $t'=0$, where 
the singularity is at half filling, it is masked by the Mott transition. 
We note that around half filling the system could become antiferromagnetic
and that close to the van Hove singularity and for large $U$ there could be 
ferromagnetism\cite{sorella}, while in our calculations we always restrict 
ourselves to the paramagnetic state.

\end{multicols}
\newcommand{\gnuplt}[2]{\resizebox{#1}{!}{\rotatebox{270}{\includegraphics{#2}}}
}
\newcommand{\sbmfplt}[1]{
 \begin{minipage}{3.3in}
 \centering
 $100\times100$ lattice, $\beta\,t=100$, $t'/t=#1$\\[-1ex]
 \gnuplt{3.3in}{SBMF100x100.#1.1d2.epsi}\\[1ex]
 \end{minipage}
}
\begin{figure*}
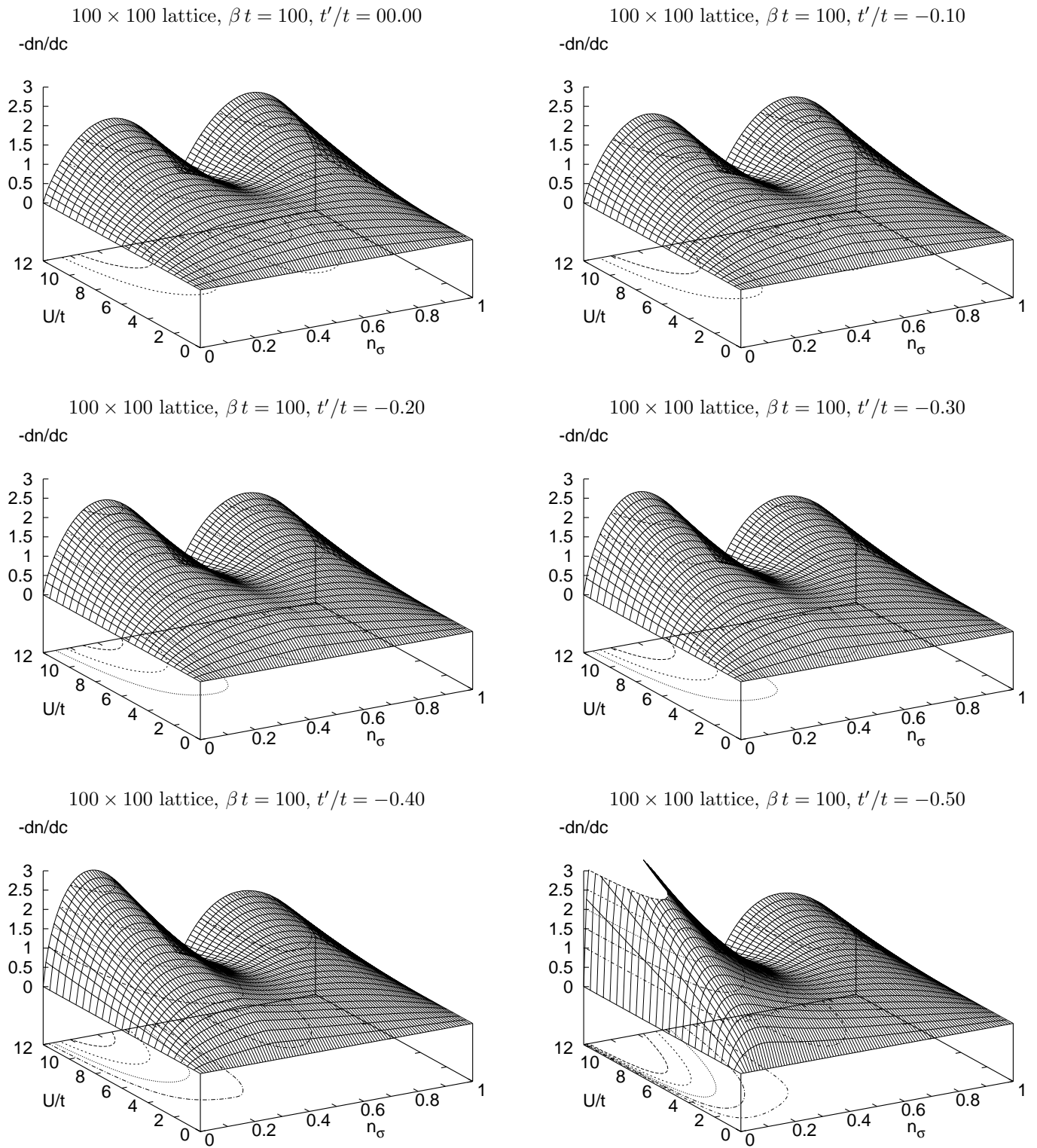

 \centering
 \sbmfplt{00.00}\hfill\sbmfplt{-0.10}\\[4ex]
 \sbmfplt{-0.20}\hfill\sbmfplt{-0.30}\\[4ex]
 \sbmfplt{-0.40}\hfill\sbmfplt{-0.50}\\[4ex]
 \caption[]{\label{sbmfresponse}
  Linear response to the perturbation by a point charge in the Hubbard model
  as a function of filling $n_\sigma$ and interaction $U$ for different
  values of the next-nearest neighbor hopping matrix element $t'$.
  The calculations use the Slave-Boson method in the mean field approximation
  and were performed for square lattices of size $100\times100$ at
  $\beta=100/t$.
  $dn$ is the electron density that is induced on a site with an infinitesimal
  test-charge $dc$, i.e.\ on a site with external potential $dc\,U$. 
  Contour lines are drawn for $-dn/dc=1,\,1.5,\,2,\ldots$, i.e.\ 
  the first contour line marks the onset of overscreening. 
  It is clear that, as the van Hove singularity (at $4t'$) is shifted
  to lower energies, the response for less than half filling gets increasingly
  stronger. For $t'=0$ the singularity is at half filling and moves towards
  smaller $n_\sigma$, until it reaches the lower band-edge for $t'/t=-1/2$.
 }
\end{figure*}
\begin{multicols}{2}

\begin{figure}
 \centerline{\resizebox{3in}{!}{\rotatebox{270}{\includegraphics{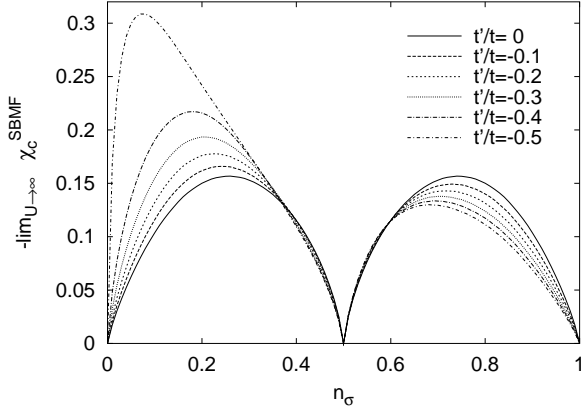}}}}
 \vspace{2ex}
 \caption[]{\label{largeUchi}
  Sucsceptibility $\chi_c=\sum_{q\ne0}\chi_q$ calculated by the Slave-Boson
  method in the limit $U\to\infty$. The calculations are for the same
  systems as in Fig.\ \ref{sbmfresponse}.
 }
\end{figure}

\subsubsection{Comparison with RPA}
\begin{figure}
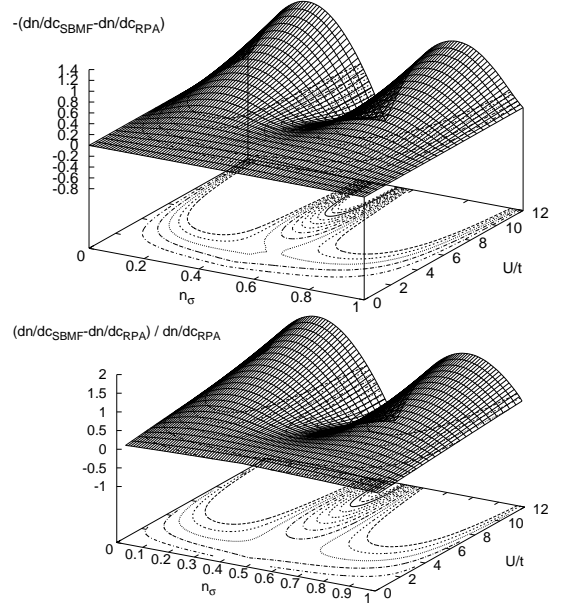

 \centerline{\resizebox{2.8in}{!}{\rotatebox{270}{\includegraphics{rpaabs-0.10.epsi}}}}
 \centerline{\resizebox{2.8in}{!}{\rotatebox{270}{\includegraphics{rparel-0.10.epsi}}}}
 \vspace{2ex}
 \caption[]{\label{RPAdiff}
  Difference between the response calculated using the Slave-Boson method
  and RPA for the Hubbard model ($t'/t=-0.1$, to avoid perfect nesting) on
  a $100\times100$ lattice at $\beta=100/t$.
  The upper plot gives the absolute difference (contour lines at -0.25, -0.2,
  -0.15, \ldots, 0.25), the lower plot the relative difference
  (contour lines at -0.5, -0.4, \ldots, 0.5).
  It shows that RPA works best close to half filling, while it quickly
  gets worse away from half filling. It is to be expected that the RPA
  fails when the overscreening sets in. But it turns out that even before that
  the RPA is already fairly bad. 
 }
\end{figure}

\begin{figure}
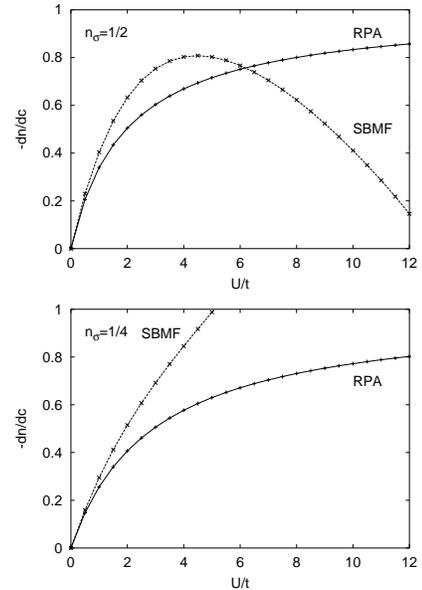

\centerline{\resizebox{2.1in}{!}{\rotatebox{270}{\includegraphics{rpahalf-0.10.epsi}}}}
\vspace{1ex}
\centerline{\resizebox{2.1in}{!}{\rotatebox{270}{\includegraphics{rpaquart-0.10.epsi}}}}
 \vspace{2ex}
 \caption[]{\label{RPAhalf}
  Comparison of the response calculated by the Slave-Boson method with RPA
  at half and quarter filling. The calculations are for the same system as 
  in Fig.\ \ref{RPAdiff} (in particular, $t'\ne0$, so there is no perfect
  nesting). 
 }
\end{figure}

The overscreening obviously has important implications for the validity
of the RPA. Since in the RPA the response can at most equal the perturbation
(perfect screening), it clearly will not work well, wherever there is 
overscreening. From the direct comparison in Fig.\ \ref{RPAdiff} we find, 
however, that it already starts to fail for much smaller values of $U$.
(The calculations are for a Hubbard model with non-zero next-nearest neighbor
hopping, to avoid perfect nesting.)
Surprisingly, RPA seems to work best close to half filling, but, of course,
only well below the Mott transition. Note that the contour lines around
$n_\sigma=1/2$ in Fig.\ \ref{RPAdiff} first mark increasing values, but,
when approaching the Mott transition, turn negative (see also Fig.\ 
\ref{RPAhalf}).
When going away from half filling, where one expects the correlated system
to become more metallic, RPA rapidly fails to give a good description of the
screening. For better visualization, we give a direct comparison of RPA and
the result of the Slave-Boson calculation in the two regimes (close to and
far from half filling) in figure \ref{RPAhalf}. For half-filling, RPA only
somewhat underestimates the response for small $U$, but fails to describe 
the eventual break-down of the screening at the Mott transition. Overall,
for $n_\sigma=1/2$ the error in $dn/dc$ is $\lesssim 0.15$ (relative error 
$\lesssim 20\%$) up to $U$ larger than $U_c/2$. This is consistent with what
was found in quantum Monte Carlo for a half-filled system with orbital 
degeneracy,\cite{screen} although it seems that the orbital degenerate system
is even slightly better described by RPA. 
Away from half-filling, RPA essentially misses to describe the steep rise 
of the response with $U$ and consequently fails to describe the screening 
even for surprisingly small values of $U$: Already for $U\gtrsim3$ the 
absolute error exceeds $0.15$ (relative error $\gtrsim 30\%$).

\section{Interpretation}
\subsection{Doping dependence}

It is clear from Fig.\ \ref{sbmfresponse} that the response strongly
depends on the electron density $n_\sigma$. Obviously, for the completely
empty ($n_\sigma=0$) or the completely filled ($n_\sigma=1$) system there is
no response at all, while for half filling ($n_\sigma=1/2$) screening 
breaks down at the Mott transition. In-between the response
has a maximum. Since this maximum, as a function of electron density, is
most pronounced for large $U$, we will consider the screening in that limit.
For concreteness, we consider a system with $n_\sigma<1/2$. Then, for large
$U$, there will only be empty or singly occupied sites. In response to an
external perturbation these electrons and holes will rearrange. Clearly, if
there are no electrons ($n_\sigma$ close to zero) or no holes ($n_\sigma$
close to one) there is no room for rearrangement, so the response will be 
very weak. In the other extreme, for quarter filling, all electrons and holes 
can participate in the screening. More generally, for the screening of 
a plane wave with nonzero wave vector, the same number of electrons and holes 
have to participate in the response, so the response is limited by the density 
of electrons/holes, whichever is smaller.
For $n_\sigma>1/2$ the argument is analogous, simply replace empty 
by singly occupied sites and singly occupied by doubly occupied sites.
We thus expect the response to be strongest around $n_\sigma=1/4$ and $3/4$,
which is roughly what is found in the calculations. There is of course also 
the prominent effect of the van Hove singularity, which greatly enhances the 
response and accounts for a shift in the position of the screening maximum, 
as well as for the strong asymmetry around half-filling.

\subsection{Overscreening}

\begin{figure}
 \centerline{\resizebox{3in}{!}{\includegraphics{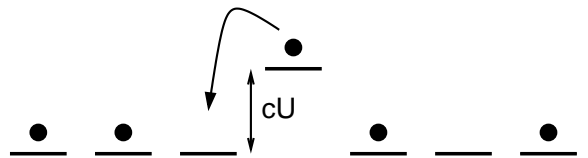}}}
 \vspace{3ex}
 \caption[]{\label{largeU}
  Screening of a point charge $c$ in the large-$U$ limit for less than
  half filling. Given a positive test charge $c$ the system will gain
  potential energy $c\,U$ by emptying the site with the test charge. Since
  there are empty sites in the system, no double occupancy needs to be created,
  so there is no cost in interaction energy. Thus vacating the site merely
  costs kinetic energy. Therefore, if $U$ is very large, an empty test-site
  is favorable: $\Delta n=\bar{n}$, independent of the perturbation $c$.
 }
\end{figure}

In order to understand the overscreening we also consider the limit of large 
$U$. Again, for a system with less than half filling ($n_\sigma<1/2$) there 
will be only singly occupied and empty sites. If the test charge is 
positive, the electrons will see a repulsive potential $c\,U$. 
Configurations with an electron on the site with the test charge will therefore
be avoided (see Fig.\ \ref{largeU}): Moving the electron from that site to some
other empty site results in a gain in potential energy $c\,U$, at no cost in 
interaction energy. Thus vacating the site with the test charge merely costs
kinetic energy. Since for large $U$ the potential energy will eventually
dominate, in the large-$U$ limit, the site with the repulsive test charge 
will be empty. This implies (i) that for a small perturbation $c$ the response
will be larger than the perturbation (overscreening) and (ii) the response
is (almost) independent of the perturbation ($\Delta n = \bar{n}$).
But this means that the response is highly nonlinear: 
while in linear response one expects $\Delta n/c \approx dn/dc$ or, 
equivalently, $\Delta n \propto c$, in the limit $U\to\infty$ we expect that 
$\Delta n$ is constant, independent of $c$! We will further explore this 
nonlinearity in the next section.

There are two questions that come to mind. First, also RPA describes 
screening as the interplay between potential and kinetic energy. So why is 
there no overscreening in RPA? The answer is that in RPA the electron-electron 
interaction is described by a mean-field: An electron interacts with the
mean electron density. So in going through the argument from above, we have
to also account for the fact that removing an electron from the site with 
the test charge increases the mean electron density on the other sites,
consequently increasing the mean-field interaction energy. Since, for the
infinite system, this increase scales the same way as the gain in potential
energy, RPA can give at most perfect screening: 
For a system with $M$ sites, neglecting the kinetic energy, we have
\begin{displaymath}
 E/U= {1\over2}(\bar{n}-\Delta n)^2
  + {M-1\over2}\left(\bar{n}+{\Delta n\over M-1}\right)^2
  + c(\bar{n}-\Delta n) ,
\end{displaymath} 
which is minimized for $\Delta n=(M-1)/M\,c$.
The second question concerns the stability of the system. 
The perturbing charge induces the system to deviate form its homogeneous
density in order to counteract the perturbation. But overscreening means
that the resulting electron density becomes even more inhomogeneous than the
perturbation, while naively one would expect that screening tends to restore 
as much as possible a homogeneous density. So does overscreening mean that 
the system prefers an inhomogeneous charge distribution, i.e.\ that it is 
instable against formation of, e.g., a charge density wave? 
The answer is, of course, no. Forming an inhomogeneous charge density costs 
at least some kinetic energy. So only when this cost is compensated
by a gain in potential energy, due to the interaction with the perturbing
charge, will the charge density become inhomogeneous. 

For an attractive perturbation ($c<0$) we can make a similar large-$U$ 
argument. Now the system will increase the electron density on the site
with the test charge. For small perturbation the system will tend to put
a single electron on the site ($\Delta n=1-\bar{n}$). But, as soon as $|c|$ is
of the order of unity or larger, it will even pay to create a double
occupancy, since the gain in potential energy ($c\,U$) exceeds the cost in 
interaction energy ($U$); then $\Delta n=2-\bar{n}$.
Systems with more than half filling can be discussed in just the same way.
In that case there is complete screening for $c<0$ by putting an double 
occupancy on the site with the test charge. For $1>c>0$ the test site will 
be singly occupied, while for $c>1$ it will even pay to empty it at the 
expense of creating an additional double occupancy.

At half filling, in the large-$U$, limit all sites will be singly
occupied and for $|c|<1$ there will be no response at all ($\Delta n=0$),
since moving charge from or to the test site would involve the creation of
a double occupancy, costing an energy $U$. 
Clearly, this Mott insulator behavior is missed by the RPA. But it is 
interesting that at half-filling the random phase approximation fails for 
the opposite reason than in the doped case. In RPA the screening always
involves a cost in interaction energy of $U\Delta n^2/2$, which, for 
half filling, underestimates the true cost $U$, while for the doped system
it is strongly overestimated. Thus at half filling RPA overestimates the
response, while for the doped system it is severely underestimated.

Finally, two points about the arguments given above might be worth mentioning. 
First, the complete-screening limits are, of course, nothing but a simple 
consequence of the Pauli principle. Second, in our arguments we have at no 
point used the spin of the electrons. Thus, as long as there are holes (for 
$n_\sigma<1/2$) or non doubly occupied sites (for ($n_\sigma>1/2$) in the 
system, there will be overscreening and nonlinear response, even if the system 
is not paramagnetic.

\subsection{Nonlinear screening}\label{nonlinsec}

As we have argued above, the response to a test charge will, in the limit
$U\to\infty$, become practically independent of the perturbation: No matter
how small the test charge $c>0$, the system (with $n_\sigma<1/2$) will respond 
in the strongest possible way (complete screening), by vacating the test site 
($n_i=0$). For $c<0$ and $n_\sigma<1/2$ the system will first respond
by putting a single electron on the test site ($n_i=1$), while for $c<1$ 
it will even occupy the site doubly ($n_i=2$). The latter, again, is
the strongest possible response --- complete screening. Thus for large $U$
the response becomes more or less independent of the perturbation, implying
a strong nonlinearity. To check this prediction, we have calculated the
response to test charges ranging from $c=-2\ldots+1$ by exact diagonalization.

\begin{figure}
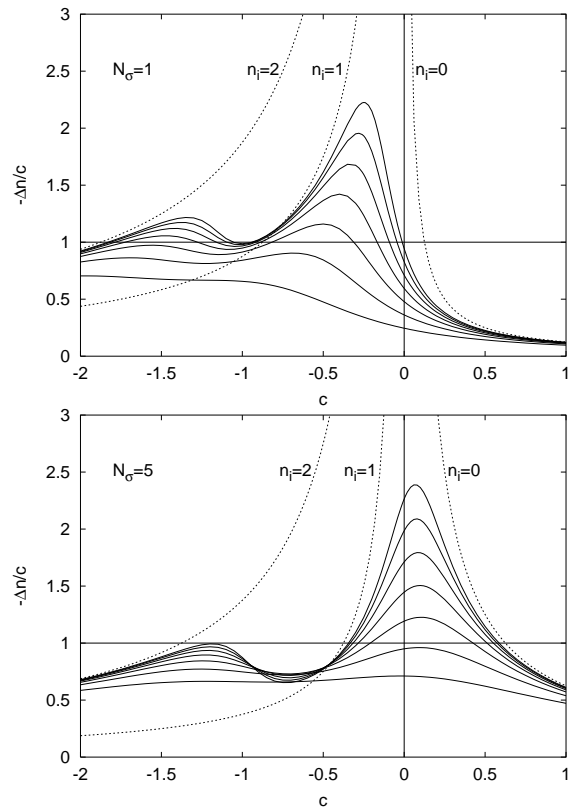

\centerline{\resizebox{2.9in}{!}{\rotatebox{270}{\includegraphics{nonlin01.epsi}}}}
\centerline{\resizebox{2.9in}{!}{\rotatebox{270}{\includegraphics{nonlin05.epsi}}}}
 \vspace{2ex}
 \caption[]{\label{nonlinearity} 
  Response to a point charge $c$ for a $4\times4$ Hubbard model ($t'=0$) with
  $N_\sigma$=1 and 5 electrons (closed shell) calculated by exact
  diagonalization.
  $\Delta n$ is the induced electron density on the test-site.
  The full lines give the results for $U=$ 4, 6, 8, 10, 12, 14, 16, with the
  response getting stronger with $U$. 
  For linear response $\Delta n/c$ is independent of $c$, so deviations of the
  curves from the horizontal indicate the nonlinearity of the response.
  The horizontal line through unity indicates perfect screening, overscreening
  above the line. 
  The limiting curves $(\bar{n}-n_i)/c$ are given by the dotted lines.
 }
\end{figure}

The results, for a Hubbard model ($t'=0$) on a $4\times4$ lattice with
$N_\sigma=1$ and 5 (closed shell systems), are shown in Fig.~\ref{nonlinearity}.
To emphasize the nonlinearity, we plot $\Delta n/c$. If the response was 
linear, this ratio would be constant (and equal to the derivative $dn/dc$). 
The deviations from the horizontal thus show the degree of nonlinearity. 
The horizontal line through unity in the plot marks perfect screening 
($\Delta n=-c$), above the response is stronger than the perturbation 
(overscreening). The dotted lines give the large-$U$ limits discussed above. 
The curves labeled $n_i=0$ and $n_i=2$ give the complete-screening limit.

Clearly, with increasing interaction $U$ the response gets stronger,
eventually showing overscreening. But at the same time also the nonlinearity 
of the response (the slope of the curve when passing through $c=0$) increases.
This is not surprising, since the curves for finite $U$ are constrained
by the response in the large-$U$ limit. Given the complete-screening limit
it is easy to see that overscreening is impossible for
$c>\bar{n}$ and $c<\bar{n}-2$.
But this implies that, the stronger the overscreening, the more nonlinear
the response has to be.

\section{Conclusion}
We have calculated the static density response for the one-band Hubbard model
using Slave-Bosons at the paramagnetic saddle-point. For finite doping we 
find overscreening, i.e.\ the response exceeds the perturbation. This can
be understood in the limit of large Hubbard interaction $U$, since strong
screening leads to a gain in potential energy which increases with $U$, 
while it merely costs kinetic energy, which is practically independent of $U$.
While overscreening thus necessarily occurs for large $U$, we found that it
actually already starts at surprisingly small $U$, in particular if the
screening is enhanced by the van Hove singularity.

The existence of overscreening might be surprising, if one assumes the
random phase approximation to give a reasonable description of the charge
response. In RPA there can be no overscreening, because the interaction is
treated in the mean-field approximation, which implies that screening always
costs interaction energy, which increases with $U$. Therefore the maximum
response possible in RPA is perfect screening, where the response just 
compensates the perturbation.
By construction RPA works well when the kinetic energy is much larger than
the interaction energy. Thus it works well for small $U$, while it necessarily
fails at the Mott transition, where it fails to describe the break-down of the
screening. Surprisingly, when doping the system away from half filling, RPA 
rapidly get worse, this time severely underestimating the response. In that
sense the doped system does not behave more metallic and RPA-like. Instead its
charge response more or less resembles what one would expect for a Wigner
crystal, i.e.\ in the opposite limit, where the interaction dominates over the 
kinetic energy. Also a Wigner crystal should show overscreening, when the 
crystal is pinned at the site of a point-charge perturbation, at the expense of
loosing the kinetic energy associated with moving the Wigner crystal as a 
whole. Clearly, the response in that case should also show the same
characteristic nonlinearity as we have found for the Hubbard model.

There have been arguments against overscreening,\cite{PinesNozieres}
based on the intuition that a system with a response exceeding 
the external perturbation would resemble an active device, in the sense of
electrical network theory, and should not be possible. More formal analyses
reveal, however, that overscreening in the charge response is indeed
possible,\cite{kirzhnits1,kirzhnits2,dolgov} and does not imply an
instability. As we have pointed out above, overscreening does not imply
that the system prefers an inhomogeneous charge density. It is only in the
presence of the perturbation that the cost in kinetic energy for forming
the inhomogeneous electron density is compensated by the gain in potential
energy from screening the perturbation.

An interesting consequence of the overscreening is that it changes the sign
of the effective interaction between external charges, i.e.\ in the case
of the Hubbard model, in the overscreening regime external charges will
experience a negative effective $U$. Moreover, we note that the overscreening
becomes the stronger the closer the van Hove singularity is moved to the lower
band edge --- in tempting analogy with the increase of the transition 
temperature in the cuprates.\cite{trends}
There are, however, two important 
observations to be made about this effective interaction. First, of course,
it only applies to external charges and not to the interaction between the 
electrons in the system. Second, the concept of an effective interaction only 
makes sense, when that interaction is to a good approximation independent of 
the charges under consideration, i.e.\ if the response is linear.

We have, however, seen that the charge response of the Hubbard model can become
strongly nonlinear. This can again be understood in terms of our large-$U$
argument. In fact, it follows, that the stronger the response, the more
nonlinear it has to become. In the extreme case the induced density becomes
independent of the perturbation. It is therefore even possible to give strict 
limits on the strength of the perturbation up to which overscreening is 
possible. This nonlinearity has to be kept in mind, when using linear response 
functions, and diagrammatic expansions for the Hubbard model.


\end{multicols}

\begin{references}
\bibitem{screen}
 E. Koch, O. Gunnarsson, and R.M. Martin, Phys. Rev. Lett. {\bf 83}, 620 (1999)
\bibitem{KotliarRuckenstein}
 G. Kotliar and A.E. Ruckenstein, Phys. Rev. Lett. {\bf 57}, 1362 (1986)
\bibitem{woelfle}
 T. Li, P. W\"olfle, and P.J. Hirschfeld, Phys. Rev. {\bf 40}, 6817 (1989)
\bibitem{rasul}
 J.W. Rasul and T. Li, J. Phys. C: Solid State Phys {\bf 21}, 5119 (1988)
\bibitem{lavagna}
 M. Lavagna, Phys. Rev. B {\bf 41}, 142 (1990)
\bibitem{woelfle2}
 P. W\"olfle and T. Li, Z. Phys. B {\bf 78}, 45 (1990)
\bibitem{li1}
 T. Li, Y.S. Sun, and P. W\"olfle, Z. Phys. B {\bf 82}, 369 (1991)
\bibitem{li2}
 T. Li, Phys. Rev. B {\bf 46}, 9301 (1992)
\bibitem{fresard}
 R. Fr\'{e}sard and W. Zimmermann, Phys. Rev. B {\bf 58}, 15288 (1998)
\bibitem{jolicoeur}
 Th. Jolicoeur and J.C. Guillou, Phys. Rev. B {\bf 44}, 2403 (1991)
\bibitem{arrigoni1}
 E. Arrigoni and G.C. Strinati, Phys. Rev. Lett. {\bf 71}, 3178 (1993)
\bibitem{arrigoni2}
 E. Arrigoni and G.C. Strinati, Phys. Rev. B {\bf 52}, 2428 (1995);
 E. Arrigoni and G.C. Strinati, Phys. Rev. B {\bf 52}, 13707 (1995).
\bibitem{seibold}
 G. Seibold, E. Siegmund, and V. Hizhnaykov, Phys. Rev. B {\bf 57}, 6937 (1998)
\bibitem{ziegler}
 W. Ziegler, H. Endres, and W. Hanke, Phys. Rev. B {\bf 58}, 4362 (1998);
 W. Ziegler, PhD-Thesis, Universit\"at W\"urzburg, 1996.
\bibitem{lilly}
 L. Lilly, A. Muramatsu, and W. Hanke, Phys. Rev. Lett. {\bf 65}, 1379 (1990);
 L. Lilly, PhD-Thesis, Universit\"at W\"urzburg, 1991.
\bibitem{preuss}
 R. Preuss, A. Muramatsu, W. von der Linden, P. Dieterich, F.F. Assaad, and
 W. Hanke, Phys. Rev. Lett. {\bf 73}, 732 (1994)
\bibitem{zimmermann}
 W. Zimmermann, R. Fr\'{e}sard, and P. W\"olfle,
 Phys. Rev. B {\bf 56}, 10097 (1997)
\bibitem{schuettler}
 H.-B. Sch\"uttler, C. Gr\"ober, H.G. Evertz, and W. Hanke, 
 cond-mat/9805133 and cond-mat/0104300
\bibitem{sorella}
 R. Hlubina, S. Sorella, and F. Guinea, Phys. Rev. Lett. {\bf 78}, 1343 (1997)
\bibitem{PinesNozieres}
 D. Pines and P. Nozi\`{e}res: {\it The Theory of Quantum Liquids},
 W.A. Benjamin, 1966. 
\bibitem{kirzhnits1}
 D.A. Kirzhnits: {\em The critical Temperature of a Superconducting System},
 in V.L. Ginzburg and D.A. Kirzhnits (Eds.),
 {\em High-Temperature Superconductivity}, Consultants Bureau, New York, 1982
\bibitem{kirzhnits2}
 D.A. Kirzhnits, Usp. Fiz. Nauk {\bf 119}, 357 (1976) 
 [Sov. Phys. Usp. {\bf 19}, 530 (1976)]
\bibitem{dolgov}
 O.V. Dolgov, D.A. Kirzhnits, and E.G. Maksimov,
 Rev. Mod. Phys. {\bf 53}, 81 (1981)
\bibitem{trends}
 E. Pavarini, I. Dasgupta, T. Saha-Dasgupta, O. Jepsen, and O.K. Andersen,
 cond-mat/0012051
\end{references}
\end{document}